\newif\ifabstract
\abstracttrue
\abstractfalse 
\newif\iffull
\ifabstract \fullfalse \else \fulltrue \fi

\documentclass[11pt]{article}
\usepackage{amsfonts}
\usepackage{amssymb}
\usepackage{amstext}
\usepackage[ruled,vlined]{algorithm2e}
\usepackage{amsmath}
\usepackage{xspace}
\usepackage{theorem}
\usepackage{graphicx}
\usepackage{url}
\usepackage{graphics}
\usepackage{colordvi}
\usepackage{colordvi}
\usepackage{subfigure}
\usepackage{lscape}
\usepackage{natbib}
\usepackage[english]{babel}
\usepackage{booktabs}
\usepackage{tabularx}

\advance \oddsidemargin by -0.8in
\newcommand{\myparskip}{3pt}
\parskip \myparskip

{\hspace*{\fill}$\Box$\par\vspace{4mm}}

\begin{document}

\title{A blockchain framework for smart mobility\footnote{\textbf{Published in the proceedings of IEEE International Smart Cities Conference 2018}}}
\author{David L\'opez\thanks{Laboratory of Innovations in Transportation (LiTrans), Ryerson University, Canada, Email: {david.lopez@ryerson.ca}} \and Bilal Farooq\thanks{Laboratory of Innovations in Transportation (LiTrans), Ryerson, Canada, Email: {bilal.farooq@ryerson.ca}}}

\begin{titlepage}
	\maketitle
	
	\thispagestyle{empty}
	
\begin{abstract}
	A blockchain framework is presented for addressing the privacy and security challenges associated with the Big Data in smart mobility. It is composed of individuals, companies, government and universities where all the participants collect, own, and control their data. Each participant shares their encrypted data to the blockchain network and can make information transactions with other participants as long as both party agrees to the transaction rules (\emph{smart contract}) issued by the owner of the data. Data ownership, transparency, auditability and access control are the core principles of the proposed blockchain for smart mobility Big Data.
\end{abstract}
	
\end{titlepage}

\section{Introduction}
Traditionally personal mobility data were solicited via small-scale surveys (1-5\% sample) and governments would take the responsibility to secure the personal information before sharing for public use. Nowadays smartphones, cellphone towers, Wi-Fi hotspots, transit counters, traffic sensors, automatic toll payment systems, among others, can passively solicit detailed mobility data of the urban population. Processing and analyzing passively as well as actively solicited data has the potential to aid governments and researchers to better understand human mobility for designing smarter, demand-driven, reliable and secure transportation systems. To fully exploit the potential of passively solicited large-scale data, privacy and security challenges need to be addressed. Passively solicited data include sensitive personal information like GPS logs or trip and activity habits, so guarding people's privacy and securing their information from untrusted parties is of utmost importance.

In recent years, cyber-security breaches have occurred all around the globe and transportation systems are not any exception. In 2015 a group of civic hackers deciphered and exposed the unstandardized bus system location data of Baltimore~\cite{rector2015}. In 2016 the San Fransisco transit was hacked to give free access to commuters for two days~\cite{wired2016}. In the same year, information of 57 million Uber customers and drivers were leaked\cite{guardian2018}. In 2018 the Ontario's regional transportation agency, Metrolinx's server was attack~\cite{ctv2018}. Blockchain technology has the potential to protect individual's personal mobility information and guard their privacy. The technology is difficult to tamper with and transactions are secure as well as transparent to all parties--including the individuals who generated the data. 

A blockchain is a distributed data structure~\cite{Christidis2016}, database~\cite{Kim2018} or shared ledger~\cite{Swan2015} that maintains a list of transaction records, which cannot be altered unless a consensus in the network is reached via proof-of-work, proof-of-stake or a byzantine fault tolerance variant. The blockchain is formed by timestamped block containing transactions and where each block is permanently linked to a previous block~\cite{Nakamoto2008}. Thus the blockchain presents a perfect solution for developing a smart network for mobility data where all the transactions are transparent (a public ledger is available to the interested parties in the network), democratic (a consensus must be reached to accept transaction) and secure (linked blocks make difficult to tamper the network). 
In this paper a blockchain framework for smart mobility data transactions is proposed. The main objective is to secure the collected data and to maintain the privacy of the individuals. The rest of the paper is organized as follows: We first introduce the background on how blockchain can solve data management problems and privacy issues in the context of smart mobility. We describe the Blockchain framework for Mobility Data Transactions. The data shared on the network as well as the rules of participation are discussed. An adversary model is presented to identify groups of attackers and how their attacks can be prevented or made difficult to succeed. At the end of the paper, a case study and concluding discussion are presented.


\section{Background}
\label{sec:brackground}

The concept of blockchain was first developed for the Bitcoin currency~\cite{Nakamoto2008}. However, in recent years the blockchain networks have gained tremendous attention due to the properties related to creating secure and private networks where no single organization is in control of the transactions and the data. It's promise is to decentralize and democratize all transactions from food to virtual currency~\cite{LeonZhao2017}. Nevertheless little research has been done outside the world of Bitcoins and its variants, only around 20\% of all research papers on blockchain focus on problems not related to cryptocurrencies~\cite{Yli-Huumo2016}. Next we present studies related to the blockchain applications in transportation and privacy.

Supply chain management is one of the main transportation applications of blockchain. The stakeholders can track their goods along the complete chain and they do not need to rely on a centralized entity for authenticity of the branded products~\cite{Crosby2016}. In combination with RFID technology~\cite{Tian2016}~\cite{Androulaki2018}~\cite{Kim2018}, the blockchain would allow the companies to track products from the creation to the delivery to the final consumer and will help them to improve their businesses by quickly identifying problems in the chain. 

Blockchain can also be used to tackle transportation supply problems. In~\cite{Sharma2017} the authors propose a blockchain network, where vehicles share their resources (fuel consumption, speed logs, space available, among others) in order to find cheap fuel stations, people for ride-sharing, or to probe good driving behavior in order to get discounts in insurance policies. The blockchain mobility consortium~\cite{Consortium2018} proposed to share and monetize the driver's information to improve network performance and to make money while driving. Applications like Arcade City~\cite{ArcadeCity2018} are proposing to share their trips in a shared mobility service, but without third parties involvement in the transaction. Shared mobility can exploit the use of blockchain to connect drivers and riders with no third parties intermediaries, however as~\cite{Stocker2016} pointed out, some issues like regulatory uncertainty, liability issues and network optimization need to be address before fully implementing blockchain for shared mobility.

Maintaining individual's privacy is currently one of the key challenges faced by various industries and researchers. Almost every part of our lives is stored on servers owned by various companies. Previously, in transportation techniques like hashing function have been used to anonymize user data ~\cite{farooq2015ubiquitous}. Public and private key encryption techniques have been used for data and communication security ~\cite{farooq2015ubiquitous}. However researchers and technologist have found that blockchain can be a potential solution to privacy problem by decentralizing information and by making the individuals the sole owners and controllers of their information. Blockchain can be used to securely share private information in: medical networks~\cite{Yue2016}, IoT networks ~\cite{Dorri2017}, access-control managers~\cite{Zyskind2015}, smart grids~\cite{ZhumabekulyAitzhan2016} and data provenance in cloud computing~\cite{Liang2017}. However, to the authors knowledge, in the literature there is no record of a generalized blockchain framework for smart mobility data transactions that can guard the privacy of individuals and protect against hacking. The framework presented in this work can be a solution to the privacy and security challenges of sharing actively as well as passively solicited large-scale smart mobility data.
\begin{figure*}
	\centering
	\centerline{\includegraphics[scale = .170]{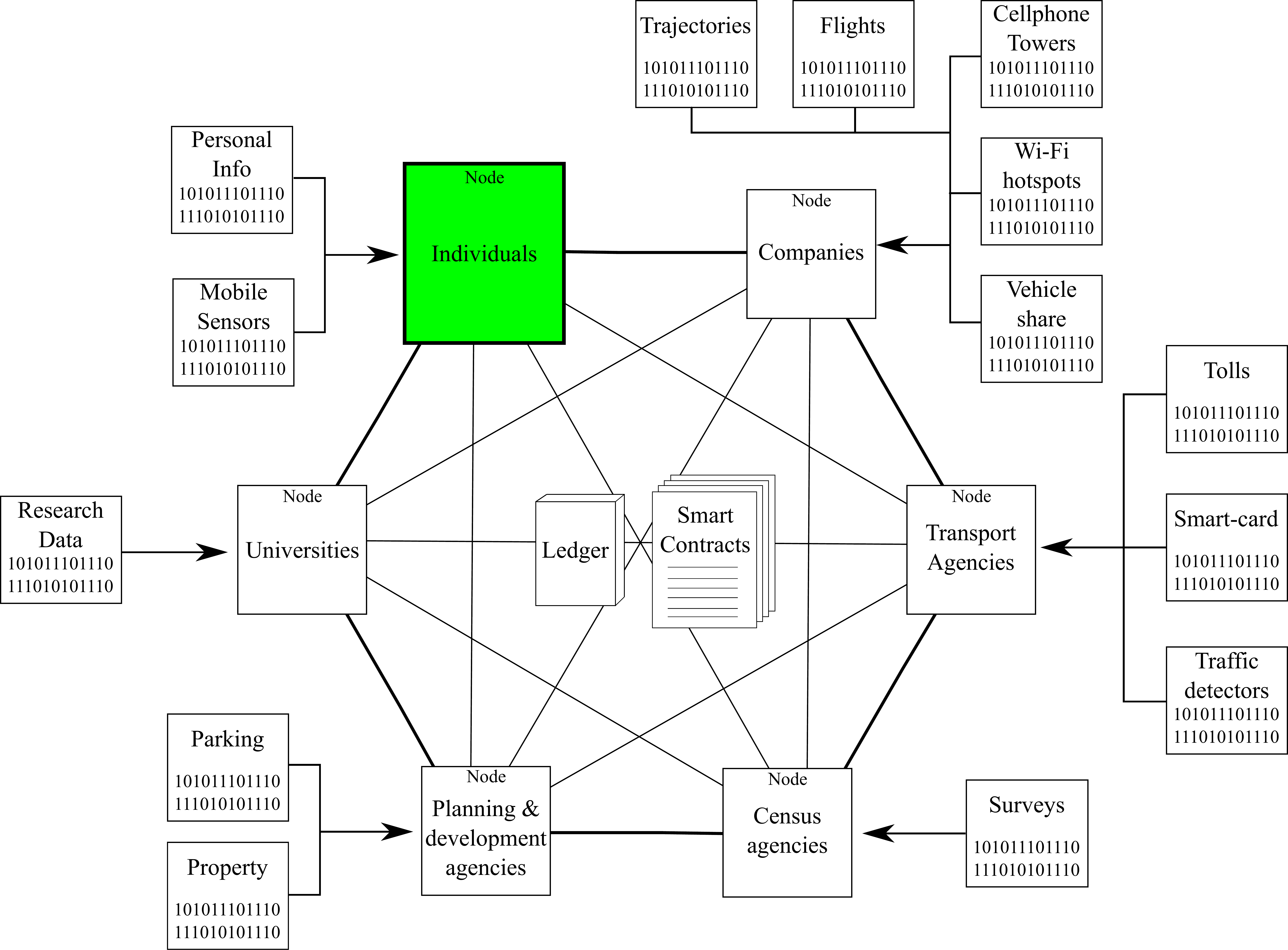}}
	\caption{Blockchain network and data collectors}
	\label{fig:diagramBlock}
\end{figure*}
\section{Conceptual framework}
\label{sec:framework}

The level of permission is the first step in the creation of a blockchain, e.g., in the Bitcoin blockchain anybody can participate and all the transactions are publicly available. Essentially there are four types of permissions~\cite{Ines2017}:
\begin{enumerate}
	\item \emph{Public closed}: Anyone can do the transactions and have access to the data. Only a restricted set of participants can be involve in consensus mechanism.
	\item \emph{Public open}: Anyone can do the transactions, have access to the data and can participate in consensus mechanisms. 
	\item \emph{Private closed}: Restricted access to transactions, have access to the data and the consensus mechanisms. Only the owner determines who can participate.
	\item \emph{Private open}: Restriction on access and who can transact. All participants can be involved in consensus mechanism.
\end{enumerate}

The main disadvantage of \emph{open} blockchains, either \emph{private} or \emph{public}, is the amount of energy necessary to reach consensus (the estimated energy consumption of Bitcoin is 100MW~\cite{Harald2017}). In the authors' opinion it will be better to consider an eco-friendlier path and to opt for a network with less consumption of energy. We propose a \emph{close} blockchain as the energy necessary to run this type of network is considerably less than an \emph{open} counterpart. 

The decision between \emph{private} or \emph{public} has to be taken in terms of access to the ledger and participation. In \emph{private} only the owner determines who can participate, giving such power to one entity or entities may lead to undemocratic process and can hurt the people's trust. The \emph{public} blockchain may fit better in the proposed framework as the participation is open to the public. However, in this type of networks the ledger is also public, so in order to protect the privacy of the individuals the personal information is encrypted and only the owners can grant permission to open selected data. Nevertheless, we believe choosing the right type of blockchain, especially one that stores personal information, should be a decision taken by all parties involved, society, government and companies need to discuss this in deep.

Fig.~\ref{fig:diagramBlock} shows the general framework of a \emph{public closed} Blockchain for Smart Mobility Data (BSMD) composed of nodes:  \emph{Individuals}, \emph{Companies}, \emph{Universities} and \emph{Government} (transport, census, planing and development agencies). The nodes collect their own data and are the sole owners of their information. All data is encrypted in the network with a personal key and transaction of data between nodes are made via \emph{Smart Contracts}. In the following paragraphs the data sources, data ownership and smart contract are explained in more details.
\subsection{Data sources}
Mobility data is constantly generated by different nodes. There are several companies producing transport information which is valuable to governments, researchers and people. For example, telecommunication companies generate data that can be used for transportation modeling, the logs of available mobile devices registered by cellphone towers or Wi-Fi hotspots can be used to monitor traffic~\cite{farooq2015ubiquitous} or to capture the individual's daily activity patterns~\cite{Phithakkitnukoon2010}. The companies can also take advantage of the blockchain to find customers or use the data generated by Government, Universities or other Companies to improve their business. It is worth to note that according to the BSMD framework companies are in control of their data so they decide to what extent they want to share information.

One of the responsibilities of the government is to collect data in order to model, manage and improve transportation networks. Information on tolls, Smart-Card, traffic detectors, surveys, parking and property can be use to find new ways for shaping our mobility. For example, toll information and traffic detectors can be used to optimize operators objective trough dynamic toll rates~\cite{Toledo2017} and Smart-Card fare data can be used to indirectly infer trip purpose~\cite{Alsger2018}.

Universities often need to collect particular data not collected by government or companies. This data is often targeted to specific porpoises like the state-preferences surveys, detailed car emissions or driver behavior. 

The biggest collector of transportation data are the individuals and their smartphones. Everyday people generate GPS points, speed, direction, mode choice, among others. In summary personal information is useful for: 
\begin{itemize}
	\item Researchers to develop cutting-edge solution for transport.
	\item Governments to plan and build better transportation systems.
	\item Companies to provide customized products or services.
\end{itemize}
However to get people's trust and to prevent the misuse of their information universities, government an companies must provide a private and secure system for the use of personal information
\subsection{Smart contracts}
\label{sec:smartContracts}
A \emph{smart contract} is a set of promises, specified in digital form, including protocols within which the parties perform on these promises~\cite{Szabo1996}. A \emph{smart contract} can be viewed as a script which defines the set of assets available to transfer and the type of transactions permitted. All \emph{smart contracts} are stored in the blockchain and have a unique address. They act as independent actors~\cite{Christidis2016} whose objective is to transact assets given a certain set of rules that involved-parties agreed upon.

In the BSMD every participant of the network defines their own \emph{smart contracts} and can only be modified by the owner of the data. All the nodes in the network can only define \emph{smart contracts} for the data they own as well as people can only define \emph{smart contracts} for the information they collect through their cellphones or by other means.

A \emph{smart contract} in the BSMD is composed of the following functions:
\begin{itemize}
	\item A function to accept connections. 
	\item A function to grant permissions to other parties to selected information.
	\item A function to revoke connections.
\end{itemize}

\begin{figure}[tb]
	\centering
	\includegraphics[scale = .2]{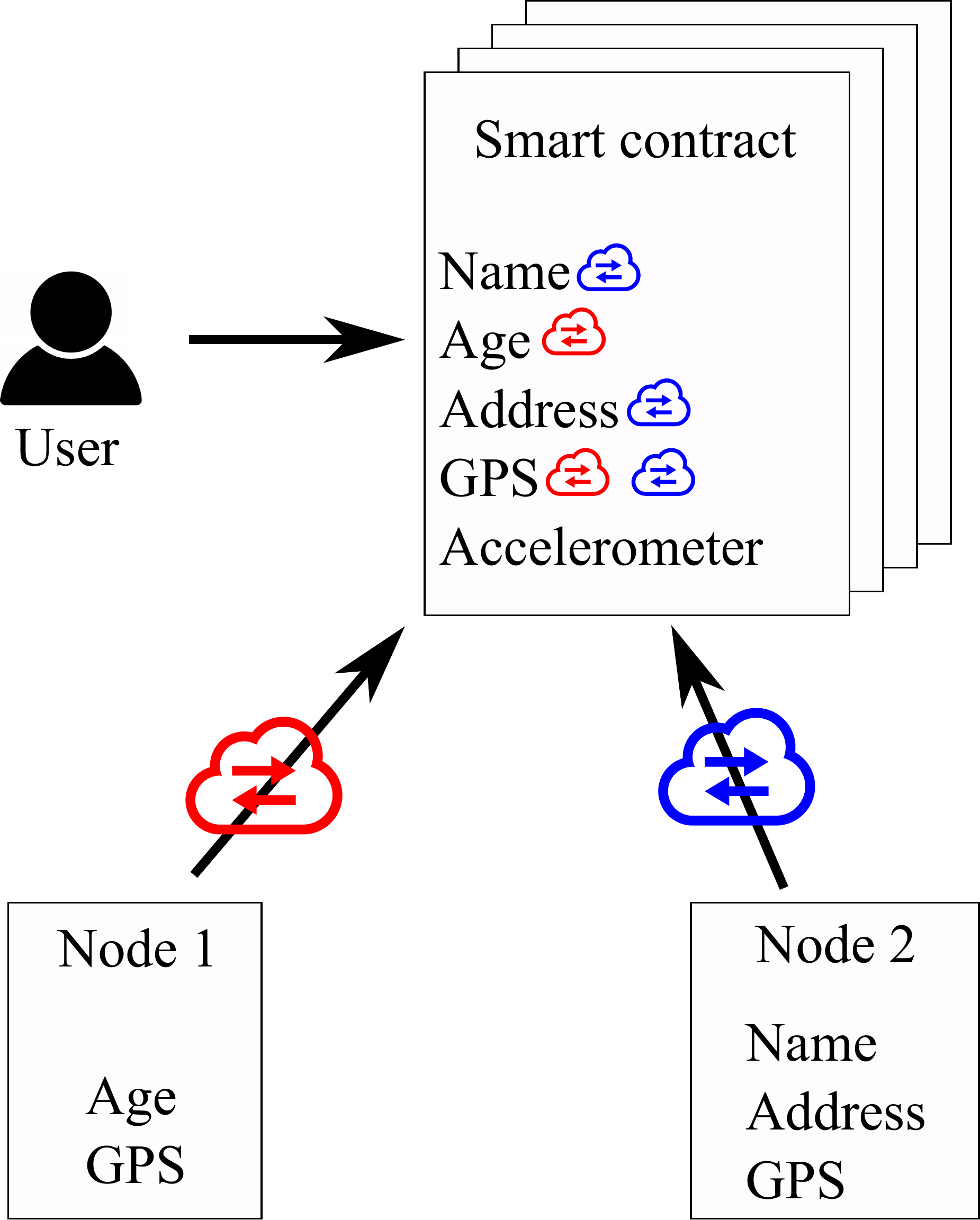}
	\caption{Smart contract and transaction}
	\label{fig:smarContract}
\end{figure}

Figure~\ref{fig:smarContract} shows a transaction using a \emph{smart contract} where an individual share selected personal data with two nodes. All data transactions are recorded in the ledger and nodes have full access to the parts of the ledger that contains their data. In this way, they can control where their information is and can detect if there is an unwanted share to untrusted nodes. 

Once a node accepts a connection it selects the information it is willing to share. As long as the connection is present in the \emph{smart contract} the receiver has access to the information. When a node requests information from another node the following steps are met:
\begin{enumerate}
	\item A receiving-node requests a connection to the \emph{smart contract}.
	\item The \emph{smart contract} confirms if the receiving-node has access to the information and accepts the connection
	\item The nodes starts sharing information in a \emph{peer-to-peer} connection.
	\item The shared data can only be accessed as long as the connection is present in the \emph{smart contract}. 
\end{enumerate}

\subsection{Users privacy}
\label{sec:UsersPrivacy}
Making the users the owners and controllers of their information is vital for generating trust so individuals feel safe when sharing their information. The proposed blockchain tackles the following privacy issues:
\begin{itemize}
	\item \emph{Data Ownership}. Nowadays companies own peoples mobility data and individuals cannot revoke access unless they opting-out. Even in that case the company may not entirely delete their information. In the BSMD the people own their data and if they want to leave the network their information is deleted from the blockchain.   
	\item \emph{Data Transparency and Auditability}. There have been many cases where companies share personal data to untrusted parties while the individuals are not aware of. In the proposed framework each individual can access parts of the ledger where their information is involved. In this manner they can know all the transaction involving them. Authorities may have full access to the ledger for auditability purposes.
	\item \emph{Fine-grained Access Control}. In case of the companies, individuals cannot choose which information they are willing to share and sometimes they cannot revoke access to specific parts of the information. With the \emph{smart contracts} individuals can manage the access to specific parts of their information.  
\end{itemize}

\subsection{Adversaries}
\begin{figure}
	\centering
	\includegraphics[scale = .235]{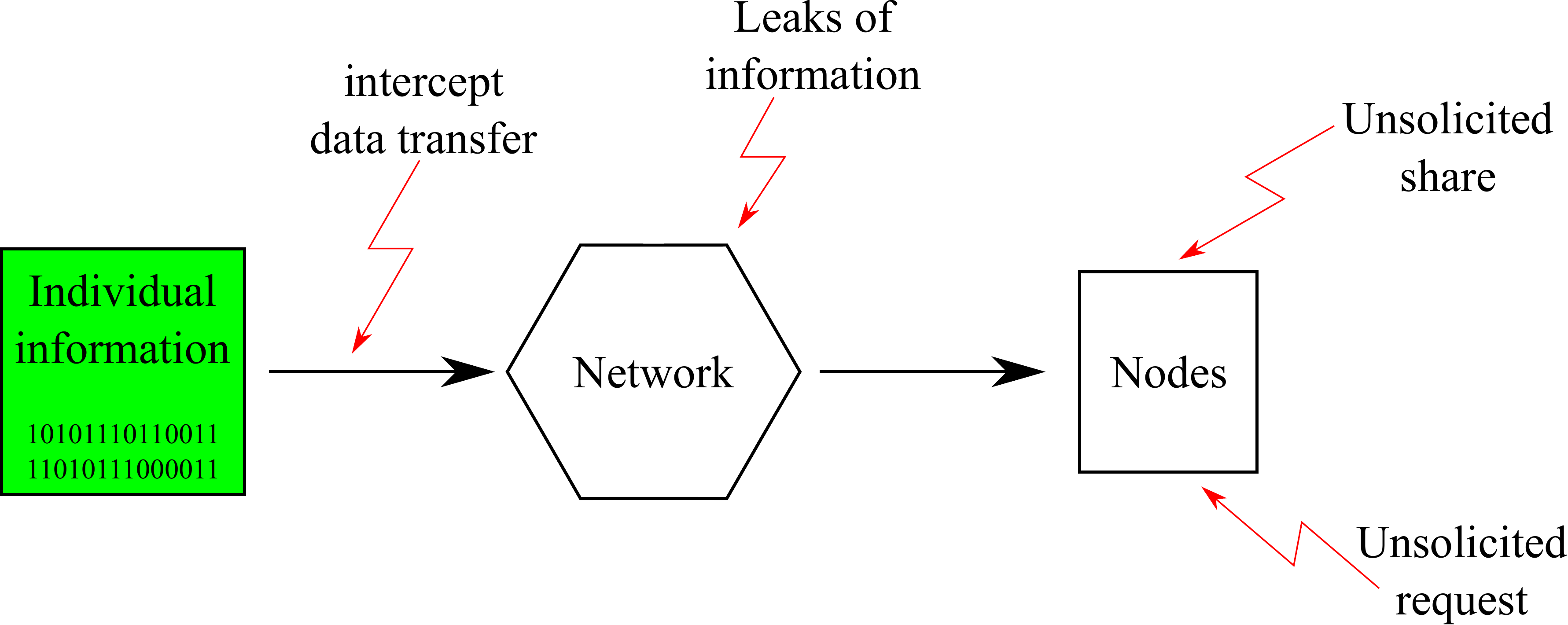}
	\caption{Adversary diagram of the blockchain}
	\label{fig:diagramAdversary}
\end{figure}

The main goal of BSMD framework is to protect the personal mobility information and secure the privacy of the people in order to fully exploit the benefits of actively or passively solicited large-scale data. To measure the level of protection in BSMD, four groups of adversaries are identified whose attacks can be prevented or hindered thanks to the use of the blockchain. Groups of adversaries may attack the network, the nodes or when the information is been transfered (see Figure~\ref{fig:diagramAdversary}). The identified groups of adversaries are:
\begin{enumerate}
	\item \emph{Data interception}: all information transactions are via a unique and secure \emph{peer-to-peer} connection, so attacking a single node may not worth the effort required to decrypt the data. 
	\item \emph{Data leaks}: all the personal information are decentralized and secured for every individual, so massive leaks on information will require huge amounts of power to hack data on a meaningful number of individuals. Also the interceptions of multiple connections at the same time will require the interception of all connections so the computing power to do this task may not be viable.
	\item \emph{Unsolicited share of information}: every mode has full access to the parts of the ledger where their information is involved so they can easily verify his information is where they want.   
	\item \emph{Unsolicited request of information}: \emph{smart contracts} let the node decide the information they want to share with specific nodes.
\end{enumerate}

\begin{figure*}
	\centering
	\includegraphics[scale = .195]{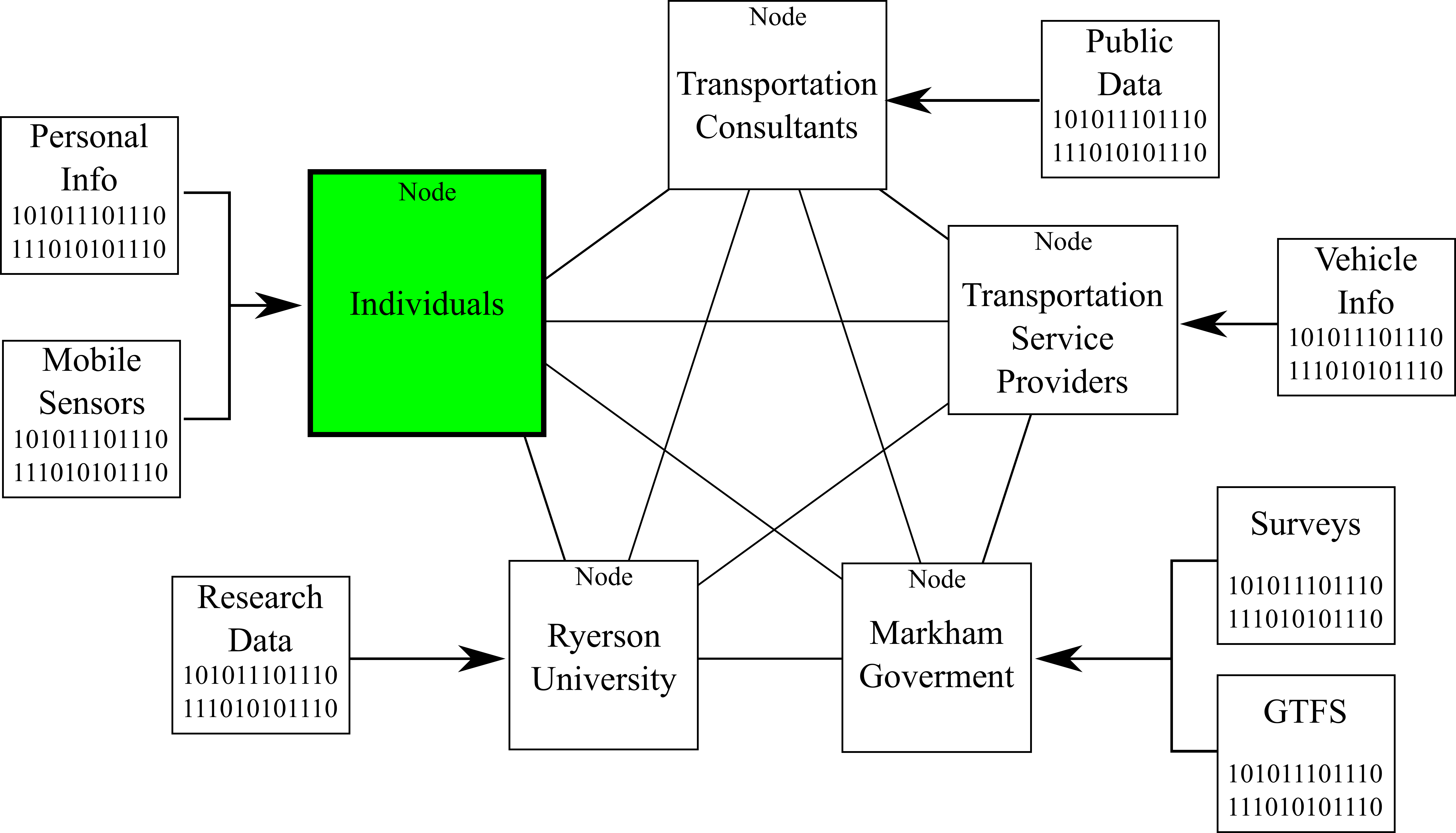}
	\caption{CarbonCount Blockchain network and data collectors}
	\label{fig:CarbonCountBlock}
\end{figure*}

\begin{figure*}[!b]
	\centering
	\includegraphics[scale = .2]{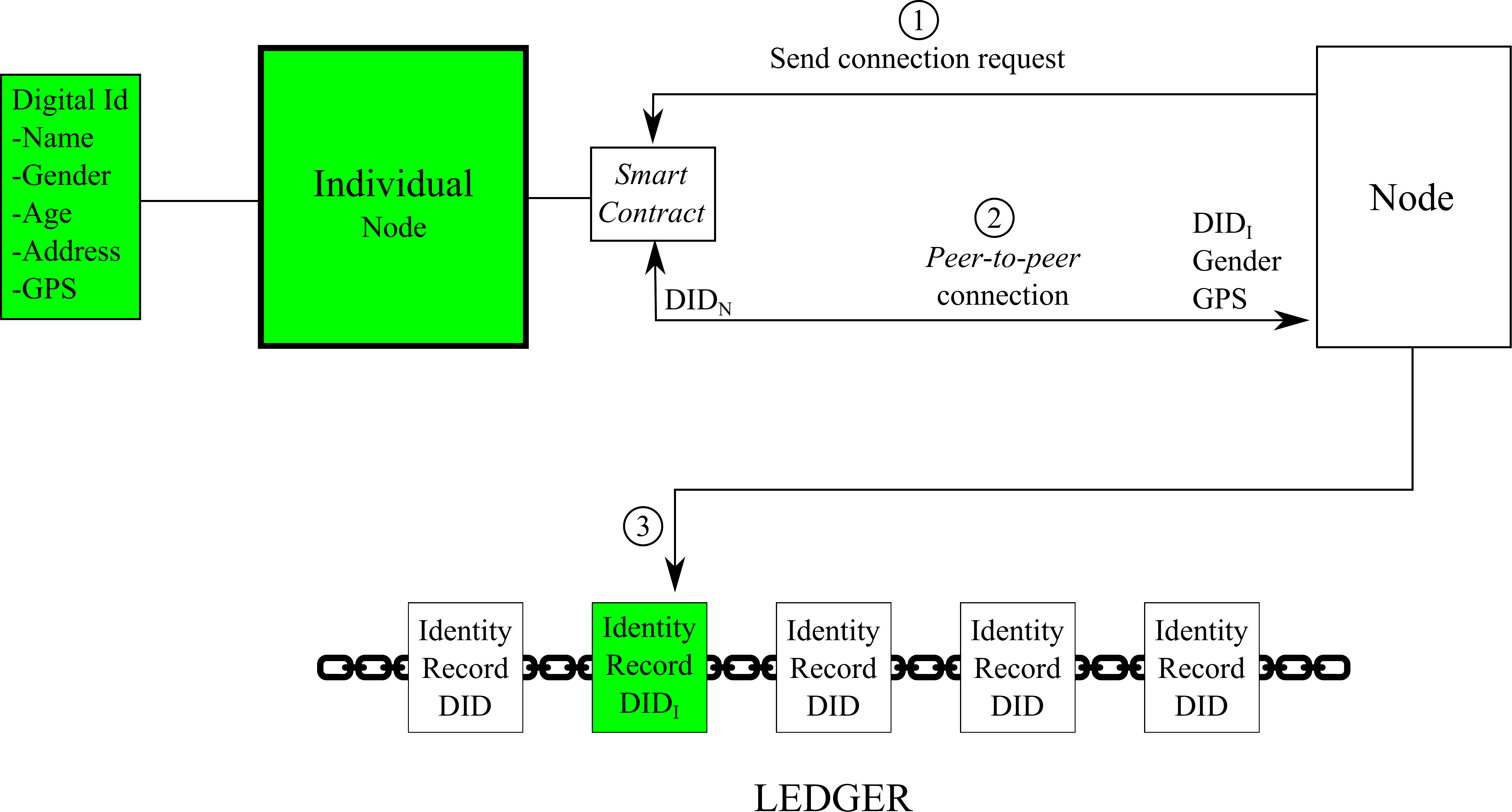}
	\caption{Transaction between two nodes in the CarbonCount blockchain}
	\label{fig:transactionIndy}
\end{figure*}

\section{Test case}

A test case is implemented for the CarbonCount~\cite{LaboratoryofInnovationsinTransportation2018} application that collects mobile sensor data (GPS, speed, direction and MAC address) and personal information from residents and share it with Markham government and Ryerson University in Canada. The goal of CarbonCount is to analyze personal mobility data to convince users to shift from cars to a cleaner modes (e.g. walk, shared-mobility, and biking) to reduce their carbon footprint. To share information between the nodes in a private and secure way the CarbonCount BSMD is build upon \emph{Hyperledger Indy}~\cite{hyperledger2018}, which is a \emph{public permisioned} blockchain for decentralized digital identities and the exchange  of information in secure \emph{peer-to-peer} connections.

The Figure~\ref{fig:CarbonCountBlock} shows the CarbonCount BSMD formed by the residents of Markham, Ryerson University, Markham government, transportation service providers and transportation consultants. On CarbonCount platform, Ryerson node will need access to the individuals origin, destination and machine inferred modes of transport to give alternative greener transport choices. Any \emph{smart contract} with Ryerson must have at least these information available if the user wants to participate in the CarbonCount app. However, if the user wants to get an alternative path that better fits their needs or suggestion in terms of activity locations/scheduling, they can opt to share more information with Ryerson. For example, if a user also share their GPS logs a tailored itinerary for reducing the emissions can be suggested. All the nodes in the network can share information for free or in exchange of services as long as the transaction are made via \emph{smart contracts}.

The mobility and personal data of the nodes are stored in a \emph{Digital Identification} that no one can revoke, co-opt or correlate without their permission. The CarbonCount BSMD ledger is public and is composed of \emph{Identity Records} associated to a Decentralized Identifiers (DID). The \emph{Identity Records} are registries of the type of transaction and the type of connected nodes, e.g., a GPS points transaction between an individual node and a government node. The DID is a sequence of bits that is generated when a connection between two nodes is made.

Figure~\ref{fig:transactionIndy} shows an example of a connection and a ledger submission in the CarbonCount BSMD. In Step $1$ the node sends a connection request to the \emph{smart contract} of the individual. In Step $2$ the individual selects which information they are willing to share and accepts the connection in the following manner:
\begin{itemize}
	\item First a secure \emph{peer-to-peer} connection is established. Transactions are direct, with no involved intermediaries to reduce the information leaks.
	\item Then, two Decentralized Identifiers (DID) are created, the individual creates a DID\textsubscript{I} and sends it to the node, while the node creates a DID\textsubscript{N} and sends it to the individual. It is worth noting that every time a connection is established a DID is created, so a single node or individual can have multiple DIDs assigned.
	\item Finally, the individual sends the requested information to the node. In the example shown in Figure~\ref{fig:transactionIndy} the individual sends their GPS logs and gender information to the node. 
\end{itemize}
Finally in Step $3$ the node submit to the ledger an \emph{identity record}  associated to the DID\textsubscript{I} so the individual can audit the ledger each time a transaction is made using their information. The validation of transactions are performed by a set of trusted nodes (Ryerson, Government and companies) running a Byzantine fault tolerant protocol.

Although other nodes or individuals have access to the ledger they cannot identify the nodes making the transaction, they can only know ``two nodes are transacting information of type $x$'', so the individual privacy is protected. Also each node has a different DID per connection making it difficult to correlate DIDs and infer the identity of the individual.

\section{Conclusions}
\label{sec:Conclusions}
The Blockchain framework for Smart Mobility Data (BSMD) transactions is presented to solve the privacy and security issues related to the sharing of passively as well as actively solicited large-scale data. Data from individuals, governments, universities and companies are distributed on the network and stored in a decentralized manner, the data transactions are recorded and must have the authorization of the owners. 

The analysis and processing of personal mobility data can improve our transportation systems and make our lives more comfortable whether in terms of going to work, building new facilities or to reduce carbon footprint. However, personal mobility data include several aspects of life that must be private, and if researchers, government and companies want to use personal data they must respect the basic human right of privacy. 

The BSMD is built on the principles of: (a) \emph{User Privacy}, multiple DIDs assigned to a single user to anonymize the information and make it difficult to correlate the data (b) \emph{Data Ownership}, each user owns a \emph{Digital Identifier} and can revoke connections; (c) \emph{Data Transparency and Auditability}, anyone can access the ledger and nodes can track all the transactions involving their information; (d) \emph{Fine-grained Access Control}, \emph{smart-contracts} define what information the node is willing to share.

This is an ongoing research to develop an operational and scalable blockchain architecture for transacting transportation data where people can trust the network as their information is private and securely distributed. The BSMD implementation will be a step forward in the use of Big Mobility Data in Smart Cities and citizen empowerment.

\bibliographystyle{plainnat}
\bibliography{bibliography}

\end{document}